\documentclass[pdflatex,sn-mathphys-num]{sn-jnl}

\usepackage{graphicx}%
\usepackage{multirow}%
\usepackage{amsmath,amssymb,amsfonts}%
\usepackage{amsthm}%
\usepackage{mathrsfs}%
\usepackage[title]{appendix}%
\usepackage{xcolor}%
\usepackage{textcomp}%
\usepackage{manyfoot}%
\usepackage{booktabs}%
\usepackage{siunitx}
\usepackage[toc,acronym]{glossaries}

\makeglossaries
\newacronym{AP}{AP}{Action Potential}
\newacronym{AFM}{AFM}{Atomic Force Microscope}
\newacronym{APTD}{APTD}{Action Potential's Time Derivative}
\newacronym{ATP}{ATP}{Adenozine Triphosphate}
\newacronym{CNS}{CNS}{Central Nervous System}
\newacronym{EM}{EM}{electromagnetic}
\newacronym{GHK}{GHK}{Goldman-Hodgkin-Katz}
\newacronym{HBP}{HBP}{Human Brain Project}
\newacronym{HH}{HH}{Hodgkin and Huxley}
\newacronym{HW}{HW}{hardware}
\newacronym{PID}{PID}{Proportional-Integral-Derivative} 
\newacronym{PSP}{PSP}{Post-Synaptic Potential}
\newacronym{AIS}{AIS}{Axon Initial Segment}
\newacronym{AI}{AI}{Artificial Intelligence}
\newacronym{AIMC}{AIMC}{Analog In-Memory Computing}
\usepackage[
]{changes}
\definechangesauthor[name=V\'egh, color=blue]{VJ}
\definechangesauthor[name=VJA, color=violet]{VJA}

\theoremstyle{thmstyleone}%
\theoremstyle{thmstyletwo}%

\theoremstyle{thmstylethree}%

\begin{document}

\title[Cross-disciplinary science for living matter]{\centering{How cross-disciplinary science}\\can describe living matter}


\author*{J\'anos V\'egh}\email{Vegh.Janos@gmail.com}

\affil*{\orgname{Kalimanos BT}, \orgaddress{ \city{Debrecen}, \postcode{4032}, \country{Hungary}}}


\abstract{
Experience shows that disciplinary science cannot describe 
life without contradictions. We show that one of the fundamental reasons
is the disciplinarity itself: the disciplines deal with
a limited set of quantities. This way some 'outlaw' quantities 
are not measured and the discipline does not 
have laws about them. All laws of science are based on approximations
and the approximations must be different for inanimate and life sciences.
Studying ions is special because ions belong simultaneously to thermodynamics and electricity, but neither of those disciplines alone can 
describe biological processes. One needs a cross-disciplinary discussion
and maybe a new scientific discipline. We provide a method for handling
the different interaction speeds characterizing the ion transport.
Electrolytes in living matter introduce 
further peculiarities with their closed volumes, internal structure, and slow processes. Their meticulous analysis led to the appropriate
approximations, leading to the correct scientific description.
As a success story, the cross-disciplinary theory of 
neuronal operation has been developed.
}



\pagebreak

\maketitle
\pagebreak
\added{
	The table of contents is present only for the comfort of reviewers
}
\tableofcontents
\pagebreak

\section{Introduction}\label{sec1}

The existence of life is still a mystery for science. As E.~Schr\"odinger formulated, "the construction [of living matter] is different from anything we have yet tested in the physical laboratory \dots it is working in a manner that cannot be reduced to the \textit{ordinary} laws of physics"~\cite{Schrodinger:1992}. Why he desperately inserted the italicised word was his firm conviction that  "\textit{not on the ground that there is any 'new force' or whatnot}, directing the behaviour of the single atoms within a living organism, but because \textit{the construction is different from anything we have yet tested in the physical laboratory}." He failed to find those laws in a disciplinary way. R.P.~Feynman was right in saying that
"the separation of fields \dots 
is merely a human convenience \dots \textit{Nature is not interested in our separations.}"~\cite{FeynmanThinking:1980} 

As Schr\"odinger and Feynman implicitly suggested, we revisit the approximations that led to classical physics (where we derived the well-known 'ordinary' laws).
We scrutinize the "construction" and its "working" in a cross-disciplinary way, using non-ordinary approximations and abstractions.
Maybe the "different construction" the living matter represents only needs different (cross-disciplinary) approximations, and they result in non-ordinary laws, which, "once they have been revealed,
will form just as integral a part of this science as the former".~\cite{Schrodinger:1992}
Maybe, then, laws based on the same first principles in a different approximation can describe living matter?

We are as general as possible when discussing the
physics of ions and electrolytes; furthermore, the physics for biology.
However, our specific goal is to establish a cross-disciplinary 
model of neuronal operation.
In section~\ref{sec:Physics-Fundamental}, we review how science
attempts to describe nature, providing sufficient accuracy, but 
requiring manageable complexity; furthermore, the dichotomies
associated with ions. 
Electrolytes are a special case within physics; furthermore, a
biological cell, with its closed volumes and special structure, represents a special case within electrolytes, as discussed in
section~\ref{sec:Physics-Biology}. Dealing with living matter requires at least the presented cross-disciplinary discussion, but maybe it 
deserves establishing a new science discipline.

Appendix~\ref{sec:Physics-SpeedHandling} is dedicated to the idea of handling different interaction speeds,
appendix~\ref{sec:statistical} explains why thermodynamics cannot be
directly applied to ions and electrolytes.
Appendix~\ref{sec:Physics-BiologyVsPhysics} compares concepts and 
conditions in biology and physics behind the different approximations,
 while appendix~\ref{sec:Physics-Speculations} shows numerical figures
(resulted from maybe somewhat naive calculations) underpinning the statements of the paper.

\section{Science methods\label{sec:Physics-Fundamental}}

Table~1 of~\cite{ThermodynamicAPDrukarch:2022} provides a concise
summary of the dichotomies disputed in the debate on modeling of nerve impulse propagation, listed below: "always both"; i.e.,
it is pointless to dispute about which one exclusively describes 
neuronal operation. Moreover, all three dichotomies (and also some others) play a role \textit{simultaneously} in biology, and they lead to 
observations differing from those in inanimate nature.
The paper above, in the context of this section, serves as a precursor to our paper and sheds light on some remarkable dichotomies.
Scrutinizing them can yield "a sound basis for unification of the physics of nerve impulses" and more.
By deeply agreeing with the fact that they have a "potential impact on our understanding of (the physical nature of) neuronal signaling", we list and uncover more dichotomies (focusing mainly on biological electrolytes); furthermore, we uncover their fundamental reasons,
partly rooted in scientific philosophy.

\subsection{Deriving abstractions\label{sec:Physics-Abstractions}}
For centuries, science developed its methods for deriving abstract concepts
by reducing the features of a real object to an abstract one that cannot be reduced further, such as mass or charge.
Ions are an exception: \textit{the ions are charge and mass simultaneously, without a further possibility of reduction}.
The consequences of this item are listed in Table~1 of paper~\cite{ThermodynamicAPDrukarch:2022}
as the "Electrical vs Mechanical" dichotomy.
Science has derived laws for the forces acting on those abstract objects, such as Newton's universal law of gravitation and Coulomb's law of electricity. Then one could apply Newton's laws of motion. Experience shows that the generated forces, regardless of their origin, can be summed, and one can apply the laws of motion using the resultant force (see below for a discussion of the resultant force and its application in biology).

\subsection{Atomicity\label{sec:Physics-Atomicity}}

Although science is aware that the apparently continuous matter in nature cannot be divided infinitely, even though it knows that the abstracted discrete
'material points' (as A.~Einstein coined) have well-defined discrete values,
it occurs rarely that both views must be applied in describing a single
phenomenon. The continuous and discrete approaches (also called macroscopic and microscopic views) seem to be independent of each other.
Connecting those views was one of the tasks performed by thermodynamics for particles with no long-range interactions. However, there is no similar discipline for electricity,
although the behavior of charge carriers is similar to that of neutral discrete 
particles. It is not evident for electrons, but it is significant for ions.
This item is listed in Table~1 of paper~\cite{ThermodynamicAPDrukarch:2022}
as the "Macroscopic vs Microscopic" dichotomy.

\subsection{Reversibility\label{sec:Physics-Reversibility}}
The third dichotomy is "Reversible vs Irreversible".
A reversible process can proceed in either direction in time.
Cellular operation seems to be cyclic (i.e., mostly reversible)
in the short run, while in the long run, spanning from birth to death, 
it is irreversible. 
Experience suggests that a mixture of reversible
and irreversible processes describes neuronal operation. 
As reviewed in~\cite{ThermodynamicAPDrukarch:2022}, in contrast with the electricity-centered disciplinary
conception \gls{HH} model~\cite{HodgkinHuxley:1952}, all other disciplinary models discussing phenomena suggest reversible operation.

Reversibility seems to be
closely related to the dichotomy "Electrical vs Mechanical"
and to the nature of the exclusively used electrical
\gls{HH} theory. In the framework of that
theory, Ohmic currents flow through resistors that irreversibly dissipate heat due to
friction, no matter in which direction ($W=I^2\times R$) the ion currents flow. 
By introducing the concept of
'delayed current' and the hypothesis that some hidden power controls the operation
of neurons by altering their conductance, physiology gave rise to the fallacy that science and life sciences are almost exclusive fields.
Furthermore, their (unintended) model provokes questions (for a review
see \cite{HH_Potential_Controversies_2017}) whether it is a model
at all, and what controversies it delivers.  
As Hodgkin wrote~\cite{HodgkinConduction:1964}: "Hill and his colleagues found~\cite{HeatProductionNeuron:1958} that an initial phase [of the action potential] was followed by one of
heat absorption. [...] a net cooling on open-circuit was totally unexpected
and has so far received no satisfactory explanation."
Since that invention, experimental evidence from the missing "leakage current"~\cite{EnergyNeuralCommunication:2021} to the conversion between
elastic and kinetic energies~\cite{MechanicalWaves:2015} to demonstrating 
the pressure-wave-like behavior of \gls{AP} witnesses that 
at least part of the phenomenon is of thermodynamic origin;
that is, it requires a cross-disciplinary explanation.
For discovering the reciprocal relations in thermodynamics~\cite{OnsagerExperimental:1959}, also in electrolytes, Lars Onsager was awarded the 1968 Nobel Prize in Chemistry.
Those relations, together with the theoretical understanding,
manifesting in the cross-disciplinary Nernst-Planck relation,
paved the way to a combined theory.
However, no thermodynamic explanation has been given so far.
Our discussion provides, in line with the experimental results,
a complex unified electrical/thermodynamic description of neuronal operation.

\subsection{Complete measurement\label{sec:Physics-Complete}}

In physics, measurement means quantifying something 
relevant to the process under study.  A 'complete' measurement
measures all relevant quantities. \textit{The different disciplines of physics restrict the measured quantities to the ones which are,
	in general, that the discipline studies.
	The remaining quantities remain outside the scope of the discipline.}
The fundamental physical quantities of mechanics are length, mass, and time, which form the basis for defining all other quantities like velocity, force, and energy in that field.
The fundamental physical quantities in thermodynamics are Temperature, Energy (Internal Energy, Heat, Work), and Entropy, which characterize systems at equilibrium and describe energy transformations; Pressure and Volume are also key variables. 
Those of electricity include electrical charge (Coulomb), the basis of all electrical phenomena; electrical current (Ampere), the rate of charge flow; voltage (Volt), the potential difference driving current; resistance (Ohm), opposition to current; power (Watt), the rate of energy transfer; and energy (Joule), the capacity to do work. 
As seen, there is little overlap of the studied quantities. 

\textit{A disciplinary study (such as the electrical and thermodynamic ones in physiology) provides an incomplete set of measured quantities}. None of the disciplines alone can describe electrolytes, since they consider different and incomplete sets of physical quantities: thermodynamics misses charge, electricity misses mass.
As discussed in connection with the Onsager relations, a disciplinary (incomplete) measurement does not discover that an unexpected change (a miracle) in the value of one of the quantities belonging to another discipline is accompanied by a change in the
value of the other quantity.. This change is clearly the case with the measurements performed in the spirit of \gls{HH}: the values the
electrical instruments provide are accompanied by the values
of mechanical/thermodynamic quantities which are not measured, partly due to the obvious 
measuring difficulties and because they are outside the scope of the discipline. 
If one attempts to reduce ion-related phenomena to a single abstraction (see the mentioned theoretical descriptions),
one experiences that some quantity (the charge or the mass)
changes in an uncontrolled way.
As Feynman R. P~\cite{FeynmanThinking:1980} told, "\textit{many of the interesting phenomena bridge the gaps between fields}". The disciplinary separation in classical science does not apply to the study of living matter.

Thermodynamics provides a framework for handling ions and determining their thermodynamic properties. However,
as good thermodynamic textbooks (including the one on the thermodynamics of the membrane~\cite{ThermalBiophysics:2007}) emphasize, \textit{thermodynamics
derives its concepts for non-interacting particles}, so one cannot expect 
its validity for ionic solutions~\cite{VeghNon-ordinaryLaws:2025}
(Boltzmann assumed that, in the absence of long-range interaction between the particles, the sizes of cells in the phase space do not change).
In addition, he required  the presence of a vast number of particles
in a homogeneous, isotropic, infinite volume,
which is typically not the case in biology.

\subsection{Instant Interaction\label{sec:Physics-Instant}}

Those dichotomies are rooted in deeper layers of science.
Notice that in the electrical abstraction, no mass is present,
so one can use the equations assuming 'instant interaction',
which in biology led to using non-physical explanations of the observations (such as 'delayed current').
In the mechanical/thermodynamic abstraction, mass is to be moved, 
making the finite-speed interaction evident.
Classical physics is based on the Newtonian idea that space and time are absolute, so everything happens simultaneously. Consequently, when nature's objects
interact, it must be instantaneous; in other words, their
interaction speed is infinitely large. Furthermore, electromagnetic
waves with the same high (logically, infinitely high) speed inform the observer. This
self-consistent abstraction enables us to provide a "nice"
mathematical description of nature in various phenomena: the classical
science.

We learned that the idea resulted
in "nice" reciprocal square dependencies, Kepler's and Coulomb's
Laws. We discussed that the macroscopic phenomenon "current" is
implemented at the microscopic level by transferring (in different
forms) discrete charges; furthermore, that solids show a macroscopic
behavior "resistance" against forwarding microscopic charges. We also learned that \textit{without charge (and, without atomic charge carriers), neither potential nor current exists}. We did not learn, however,
that also thermodynamic forces can move the ions,
given that they have inseparable mass.

The low speed of ions in electrolytes introduces problems.
The same physical phenomenon,
the interaction (or movement) of ions is described using an 'infinite' electrical speed and 
a million times less mass propagation speed, respectively, which leads
to an unresolvable discrepancy, given that physics is not prepared to handle
different speeds in the same interaction event~\cite{VeghNon-ordinaryLaws:2025}.
(Historically, Onsager's work in 1931 referred to thermoelectricity and transport phenomena in electrolytes, thus connecting experimentally two separate disciplines.)

\subsection{Locality and finite resources\label{sec:Physics-Resources}}

The closed volume of the biological objects is also
a problem of finite resources. The material transport represents simultaneously mass and charge,
so the transport itself gradually changes the gradients. 
This process keeps the entire volume of the electrolyte in (more or less intense) continuous change, which makes life possible.
The more distant parts of the biological cell will see any change in the local values of the state variables with a delay. In biology, we cannot use the
fundamental assumption of physics that, although a field acts on the 
ion in question, the ion does not affect the field (that the other ions generate).  The transferred ions decrease the field in the volume segment they departed from
and increase it in the volume segment where they arrived. Due to the 
field-dependent speed within the electrolyte, we must consider the autonomous
change between the microvolumes where the ion traverses. 
Furthermore, biological objects inside the cell
can absorb ions and charge up. With their potential, they alter local gradients, accelerating or decelerating ions.  
Not to mention that biological objects can be active in the sense that 
(depending on the environmental conditions) they can let ions from one separated volume part into the other, thus changing the number of 
charge carriers in the volume under study.

\subsection{Mathematical issue\label{sec:Physics-Mathematics}}

The interdependent behavior of charge and mass has an interesting
mathematical consequence as well. 
The definition of a partial derivative of a function of several variables is its derivative with respect to one of those variables,
\textit{with the others held constant}. In the case of ions, changing the function with respect to mass or charge means simultaneously changing also the other; that is, the partial derivatives depend on each other.
In other words, one cannot calculate the partial derivatives; only the total derivative.
As a consequence, equations that use the partial derivatives of concentration and electrical potential are either incorrect or approximations. Classical mathematics cannot be applied.
Consequently, in the case of ions and electrolytes, one must not use the well-established concepts (enthalpy, entropy, etc) of thermodynamics in unchanged form. (BTW: the discrete nature of charged particles 
leaves the question open, under which conditions remains the 
ion current differentiable.)

\section{Physics for biology\label{sec:Physics-Biology}}

The above items 
are what E.~Schr\"odinger coined as "\textit{the construction is different from anything we have yet tested in the physical laboratory}"~\cite{Schrodinger:1992}. Consequently, measurements must be designed and carried out with care; the routine
methods used for measuring objects from inanimate nature
cannot surely be applied to living objects without changing them.

\subsection{Equilibrium state\label{sec:Physics-SteadyState}}

When in the volume an
invasion happens, electrical potential, pressure, temperature,
or concentration changes locally; dynamic changes begin to restore
the balanced steady state. When the invasion persists, the system
finds another equilibrium state. 
An observer experiences that changing
one macroscopic parameter of the system causes an unexpected (and
unexplainable) local change in another macroscopic parameter(s).
Experimentally, the microscopic world maps the change from the world
of electrical abstraction to the world of thermodynamic abstraction
and vice versa; a cross-disciplinary experience occurs. Theoretically, we can do the exact mapping of macroscopic
electrical and thermodynamic parameters using microscopic universal constants. 
The discussions, for example, the one leading to
Eq.~(11.30) in~\cite{KochBiophysics:1999}, assume that the 
interaction speeds are identical and that the partial derivatives 
can be interpreted for ions. Both assumptions are wrong. However,
the claim that "while diffusion is like a hopping flea, electrodiffusion is
like a flee that is hopping in a breeze"~\cite{KochBiophysics:1999} (attributed to Hodgkin) is true; see also our Eq.~(\ref{eq:PhysicsGradientRatio}).

We can describe the equilibrium
state (the mutual dependence of the \emph{spatial gradients} of the
electrical and thermodynamic fields on each other) using the famous Nernst-Planck transport equation 

\begin{equation}
	\frac{\partial c}{\partial t} = \nabla \cdot
	\biggl[ \underbrace{D\nabla c}_{Diffusion} -\underbrace{\mathbf{v}c}_{Advection} + \underbrace{\frac{Dqz}{k_{B}T}\mathbf{E}}_{Electromigration} \biggr] 
	\label{eq:NernstPlanckTransport}	
\end{equation}
Unfortunately,  Eq.~(\ref{eq:NernstPlanckTransport}) also comprises the issue specific for ions: it
calculates the not-interpreted partial derivative without taking into account that at least the advection term modifies the local electrical gradient by changing the concentration.
Assuming that the concentration is at equilibrium ($\frac{\partial c}{\partial t}$ = 0; even if its calculation method and its value are questionable), without an external electrical field ($\mathbf{E}$ = 0),
the change due to diffusion (in a statistical sense) is zero, and the flow velocity is zero ($\mathbf{v}=0$),
results in the Nernst-Planck equation in one dimension
\begin{equation}
	\frac{d}{dx}V_{m}(x)=-\frac{RT}{q*F}\frac{1}{C_{k}(x)}\frac{d}{dx}C_{k}(x)\label{eq:NernstPlanck}
\end{equation}
In good textbooks (see, for example,~\cite{KochBiophysics:1999},
Eq (11.28)), the derivation of the equations is exhaustively detailed. In the
equation, $x$ is the spatial variable across the direction of the
changed invasion parameter, $R$ is the gas constant, $F$ is the
Faraday's constant, $T$ is the temperature, $q$ the valence
of the ion, ${V_m(x)}$ the potential, and ${C_k(x)}$ the concentration
of the chemical ion. In simple words, it states that the change in the
concentration of ions creates a change in the electrical field (and
vice versa), and in a stationary state, they remain unchanged. However,
\textit{in classical science, there is no way to take into account the
	field's propagation speed}. We must call attention to a nuance: to keep balance, the concentration and the potential must change in opposite directions. However, since they are implemented by ions, moving an ion's
mass changes the ion's charge in the same direction.

It is one of the rare cases in which the starting point was wrong, but the conclusion was correct.
In Eq~(\ref{eq:NernstPlanckTransport}),  
an identical speed for all interactions was assumed. Although the equation is not really applicable for describing the transport of ions (in its practical applications, the identical speed
was calculated as a "mean-field", where the "mean" stands for some average
of interaction speeds differing by several orders of magnitude), in an equilibrium state, the actual value of both
interaction speeds is zero, so they have the same value; furthermore, no advection takes place. This way, the wrong partial derivative is irrelevant.

In 'ordinary' physics, where the charge and mass are independent,
changing one side changes the other in the opposite direction.
In 'non-ordinary' physics, the two differentials must change in the same direction, given that ions' charge and mass cannot be separated.
Correspondingly, when initiating an \gls{AP} in a neuron, the rush-in of ions
increases both concentration and potential simultaneously. 
Moreover, to restore equilibrium, the gradients of the two macroscopic parameters must have the same sign, since they refer to the same ion.
The thermodynamic force acting on an ion is added to the electrical force.
The two are inseparable; the magnitude of the force is 'falsified'.

Eq.~(\ref{eq:NernstPlanck}) is the derivative of the equation
\begin{equation}
	V_{m} =\frac{RT}{q*F}\ln{\biggl(\frac{C_{k}^{ext}}{C_{k}^{int}}\biggr)}\label{eq:Nernst1}
\end{equation}

\noindent known as \textit{Nernst law}.  In other words, Eq.(\ref{eq:NernstPlanck}) and Eq.(\ref{eq:Nernst1}) are the differential and integral formulations, respectively, of the same knowledge. (Eq.~(\ref{eq:Nernst1}) results in opposite signs according to the 'ordinary' and 'non-ordinary' laws of physics. Experience shows that the 'non-ordinary' laws result in the correct sign.)

That means, of course, that the Nernst law is valid only in a system 
at equilibrium with no material flow (it cannot be interpreted for transient states, such as generating \gls{AP} in a cell).
The limits of the integration are chosen arbitrarily (by choosing the reference concentration $C_{k}^{int}$).
Hence, the derived potential values inherently include an additive term (a potential difference), so one must not compare them directly if they use different reference potentials $C_{k}^{int}$; see reversal potential values.
Furthermore, it is nonsense to combine the quantities from the intracellular and extracellular sides additively,
whether concentrations, or mobilities, or Nernst voltages, as it happens in the \gls{GHK} equation, see~\cite{VeghMembranePotential:2025}.


In the steady state, in contrast with the case of an ion in an infinite space, some other forces also contribute to
the mentioned ones. To discuss how nature restores the steady state when a microscopic change occurs in a
balanced state of a biological solution, we write the well-known Nernst-Planck
equation (see Eq.(\ref{eq:NernstPlanck})) in a slightly extended form:
\begin{align}
	\underbrace{\bigl(-F_{constr}(z)\bigr)}_{Constraint} &= \underbrace{ e_{el}*E_{Gap}^{Total}(c,\Delta z)}_{Electrical} +\underbrace{\mathbf{e_{el}*E_{thermal}^{C_k}(d)}}_{\textbf{Thermodynamic}}\\
	&\underbrace{\bigl(+F_{Transp}(z)\bigr)}_{Transport}\underbrace{\bigl(+F_{Invasion}(z)\bigr)}_{External} 
	\label{eq:NernstPlanckExtended}
\end{align}
We multiplied the usual two terms by the elementary charge, so they are expressed as forces, plus we added an external force (its role is discussed below). Furthermore, changing one force triggers a corresponding counterforce and/or causes the ion to move in a viscous fluid. 
All that means when describing an ionic transfer process, \textit{we must not separate the electrical current from the mass transfer}: they happen simultaneously and mutually trigger each other. Notice that the thermodynamic term is ion-specific while the electrical term is not. To be entirely balanced, the system must be balanced to all chemical elements. In this way, changing one concentration implicitly changes all other concentrations and the electrical field.
That is, the speed of material transport gradually changes
as the steady state is approached; furthermore, it is by orders of magnitude smaller than the speed of the interactions.
When neuroscience teaches that "pumps that maintain
ion gradients \dots transport ions against their
electrical and chemical gradients"~\cite{PrinciplesNeuralScience:2013}, page 101, one shall ask what kind of force acts on the ion, keeping in mind
E.~Schr\"odinger's opinion that no "'new force' or whatnot"~\cite{Schrodinger:1992} affects ions' motion.  

In balanced states, no transport occurs, so the transport force cancels out, and the mobility has no role. The counterforce adapts to the situation.
That force may be a mechanical one:
the ions sitting on the surface of the membrane press the
surface due to the attractive force of ions, and the membrane mechanically provides the needed 
counterforce. 
If the boundary of the segments is not freely penetrable, the counterforce equals the difference between 
those two forces. In this way, no resultant force acts on the ions; the two gradients persist. When ions can move freely between segments, they will continue moving until they produce a concentration gradient (a thermodynamic force) for the given ion that counterbalances the electrical gradient (the electrical force), at which point transport stops. A transport force is needed to reach the Stokes-Einstein speed in a viscous fluid.

An important effect is that \textit{the counterforce combines two disciplines}.
In neurons, when $Na^+$ ions rush into the membrane at the beginning,
they produce, simultaneously, a huge electrical and thermodynamic gradient.
The repulsion among the ions creates a huge mechanical pressure increase (an impact force)
that presses the elastic membrane. The counterforce starts a mechanical 
shock wave (soliton)~\cite{SolitonPropagation:2005}, that is simultaneously
an electrical shock wave~\cite{VeghMembranePotential:2025} that is measured as \gls{AP}. The mechanical shock wave generates a damped oscillation in the fluid.
The electrical impact force produces a huge electrical gradient, and the rushed-in charge discharges through the $RC$ circuit. The charge carriers have both mass and charge, so that a voltmeter can observe the superposition of the two effects as \gls{AP}. Only using a cross-disciplinary discussion (non-ordinary laws) enables understanding the neuronal operation, i.e., the two phenomena simultaneously. 

In closed volumes, see Schr\"odinger's points on the specific aspects of life, one must consider further effects.
The electrostatic forces of the ions exerted on
each other, and the wall exerts a counterforce 
on the ions in the microscopic view, and on the fluid in the macroscopic view, which manifests as a mechanical force and pressure.
This aspect is not considered in the general transport equation~(\ref{eq:NernstPlanckTransport}); one must add one more term
when calculating mass transport.
If the volume is not hermetically closed (say, the ion channels in the 
wall of the membrane represent a "hole" for the ions), the combined 
electrostatic plus mechanical pressure can provide a driving force (instead of hypothesizing a magic lipid mechanism)
for producing a mass transport. If the change in concentration and potential happens suddenly (see the rush-in of $Na^+$ ions at the
beginning of an \gls{AP} in neurons), that "explosion" creates an "impact force" exerted on the membrane.
Furthermore, the counterforce starts a pressure wave and other mechanical changes, as discussed in~\cite{MechanicalPropertiesNerves:2025}. \textit{The electrical and thermodynamic phenomena are just two sides of the same coin} as discussed in~\cite{VeghMembranePotential:2025}. Fortunately, 
the overwhelming majority of the energy of the excitation
is stored as elastic energy of the membrane, and the pressure is proportional to the measurable voltage, so describing the electrical behavior is sufficient to provide a sufficiently precise description
of the processes in the membrane, but one has to keep in mind
that the force is composed of electrical and thermodynamic components.

It was a colossal mistake, which takes its origin in the disciplinary view of electricity, to introduce equivalent circuits with their fixed-value voltage generators. It forces one to assume that the 
conductances of the players (membrane, synapses, \gls{AIS}) 
change 
without any reason, and prevents understanding how the competition between thermodynamic and electrical processes governs neuronal operation.
It leads, among others, to attributing conductance change to membranes, which are simple isolators with no charge carriers that can implement
charge transfer; in this way
attributing the change of an electrical entity to the biological material.
This assumption neglects the driving force that the mentioned forces provide. 
Instead, it attributes the magic ability to the ion channels that they
can change their transmission ability as the actual situation requires.

\subsection{Laws of motion\label{sec:Physics-LawsOfMotion}}

Scientific laws about the separate interactions of masses and charges are based on abstractions that require approximations and omissions.
While we understand that the speeds of electrical and gravitational
interactions are finite, they are so large that we can use the 'instant interaction' approximation
in classical physics. One effect of the first particle
reaches the second particle simultaneously with the other effect,
leading to the absence of a time-dependent term in the mathematical
formulation. However, this is not the case in electrodiffusion, where
the mass transfer is significantly slower than the transfer speed
of the electromagnetic field. 
\textit{Science uses the notion 'instant' in the sense that one interaction
	is much faster than the process under study; we consider the faster interaction as instant.}

From a physical point of view, ionic solutions are confined to a well-defined volume with no interaction with the rest of the world.
One must adapt the relevant laws to the case of finite resources and account for the counterforces exerted by the boundaries.
Furthermore, internal processes result in slow transport,
which, together with the biological "construction", leads to local concentration gradients.
At a microscopic level, on the one hand, we use the abstraction that they consist of charge-less, size-less, simple balls with mass, have thermal (kinetic) energy, and collide with each other, as thermodynamics excellently describes.
On the other hand, we use another abstraction: massless, sizeless charged points with mutual repulsion.
All respective laws contain only one of those abstracted features.
At a macroscopic level, we use the abstraction that the respective volume is filled with a continuous medium with 
macroscopic parameters such as temperature, pressure, concentration, and potential.

One can parallelize the description of how ions change their positions using Newton's laws of motion, relating an object's motion to the forces acting on it. The first and third laws are \textit{static}, and the second is \textit{dynamic}. We can translate the first law into ions: without an external invasion, their volume at rest will remain at rest (apart from diffusion). The third law for ions' volume essentially states that, in a resting state, at every point, the electrical and thermodynamic forces are equal; the Nernst-Planck electrodiffusion equation expresses this.
The second law, for mechanics, expresses the time course of the object: \textit{the position's time derivative}.
Notice that, in this case, we make \textit{one abstraction}: the object (the carrier) has \textit{one attribute}, its mass.
(Recall how important it was for the special theory of relativity that the \textit{accelerated mass} and the \textit{gravitational mass} were identical.) 

For ions, we have \textit{two abstractions}, the attributes \textit{'charge' and 'mass'}, and two forces acting on these attributes, which science classified as belonging to different disciplines.
We cannot easily express how electrical and thermodynamic forces will change an object's position because they act differently on different attributes.
No time derivatives are known, only position derivatives. (Recall section~\ref{sec:Physics-Mathematics} about partial derivatives: derivations such as Eq. (11.30) of~\cite{KochBiophysics:1999} are wrong.)
Due to this hiatus, physics (and consequently, physiology) cannot describe the electrochemical processes: \textit{the second law of motion for electrodiffusion is missing}.
As a consequence of the instant interaction, \textit{classical science has no mechanism for handling the case
	when two different force fields (gradients) having different propagation speeds act on an object
	and two different abstractions (charge and mass)},
belonging to different science disciplines) translate the force into acceleration. Here we use Boltzmann's idea: we derive the particles' speed
from the continuum mass's speed.

When describing processes (i.e., dynamical systems), we must have
one or more equations of motion:
how the time gradients of the fundamental entities change as they function. In classical
science, we have only one fundamental entity: the position, and the driving
forces depend only on the position. The (Newtonian) laws of motion do not depend on a time gradient.
In the Einsteinian world, speed explicitly appears when describing
the interrelation of fundamental entities mass, position, and time.

\textit{In our 'non-ordinary' science, we have a double-speed interaction} abstraction; correspondingly,
we have \textit{two} laws of motion. 
In our laws of motion (see Eq.(\ref{eq:Nernst-dVdt}) and Eq(\ref{eq:Nernst-dCdt})) we also have an explicit speed dependence 
in describing the interrelation of concentration and potential.
Actually, we are thinking in two entities (concentration and potential), but both of them are parametrized by position $x$.
In line with the Einsteinian case, the time shines up, and 
again, the speed connects those entities.
\textit{However, the different interaction speeds act on the coordinates
	differently; that is, the effects of changed entities cannot
	be separated. This non-separability is why we need 'non-ordinary' laws.}
(Given that our thermodynamic speed is always by orders of magnitude lower
than the electrical (limiting) speed, we neglect transforming the time.)

In all cases, the law has the form of a differential equation;
i.e., we can derive the fundamental entities by integration.
\textit{In 'ordinary' science, we have a single-speed interaction} abstraction;
so we have one law of motion; an analytical solution is possible. 
In a non-ordinary case, we have a dual-speed interaction, and a numerical solution is likely the only option.

\subsection{Time derivatives\label{sec:Nernst-time-derivatives}}

Eq.(\ref{eq:NernstPlanck}) describes a stationary state with no ionic
movement. Deriving a time course (time derivatives) from the position
derivatives is impossible in a strict mathematical sense.
However,
we can provide it using physical principles.
We consider the electrical ion current represented by a viscous charged
fluid~\cite{ViscousChargedFluids:2014}. As expected, selecting the
speed (aka calculating the appropriate value of the local macroscopic speed,
see Eq.(\ref{eq:StokesEinsteinSpeed2}))
plays a key role, especially since we are at the boundaries of physics
abstractions: we are mixing microscopic and macroscopic
concepts.

In classical
physics, because of the lack of time-dependent terms in the expressions,
the changes are described by position-dependent terms (position
	derivatives), both in the case of electromagnetic and electrodiffusional
interactions. In classical ('instant interaction') science, the
time derivatives are either not interpreted or can be derived through
the externally derived joint interaction speed. As explained, one can
extend the idea to enormously different speeds and derive time derivatives by
considering the faster interaction as an instant.

In timeless classical physics, there is no explicit dependence on
time: everything happens simultaneously. In a resting state, the
Maxwell's equations follow from the conservation of energy.
One form of energy transforms into another form, and the system arrives
in another balanced state. The carriers of force fields are continuous,
so one can calculate and make infinitesimal changes in the driving
forces; they do not change the system's energy. If one gradient changes,
the other automatically (per definitionem) changes in the opposite direction. In other
words, the driving forces are permanently balanced, the magnetic and
electrical forces act instantly ("simultaneously"), and are
always of opposite sign. A time derivative cannot be interpreted:
everything happens at the same time. In other words, at the same space-time
(in the classical interpretation, the time is the same at any point).

In an electrodiffusional process, we start with the same
point of view. We assume that the thermodynamic and electrical driving
forces are equal in equilibrium. That assumption results
in the Nernst-Planck equation. On the one side, we use a macroscopic parameter,
the potential. On the other side, we use another macroscopic
parameter, the concentration. The equation bridges those
macroscopic parameters by using universal constants from the microscopic world.

At first glance, the case is similar to that of electrical and magnetic fields.
However, we cannot
make infinitesimally small changes in the gradient since the carrier
of the force fields is "atomic". Furthermore, moving it infinitesimally
(changing only its position coordinates), the changes in the electrical
and thermodynamic gradients do not result in a new balanced state.
The effect of ions' \emph{charge} has an immediate effect on the volume,
but ion's \emph{mass} has a delayed effect. The infinitesimally small
change in the position results in an infinitesimally small increase
in the energy of the system, given that moving a carrier changes the
potential and the concentration in the same direction, since we did not
consider that the
time coordinate changes as well.
In the Newtonian world, everything
happens at the same time, so we cannot handle instant and finite interaction
speeds simultaneously.
Eq.~(\ref{eq:Nernst1}) contradicts energy conservation.
The infinitesimally small energy change disappears
only when the effect of the slower interaction reaches the other carriers in the
volume. \textit{When the interaction speeds differ, energy conservation is valid only if one uses space-time.}

Fortunately, we can derive the infinitely small change where
the time and space (position) coordinates are connected; essentially, in the same way as in
the special theory of relativity. Let us assume that the gradients
act on the mass and the charge, but the ion's effects on the gradients
are negligible. According to the principle of relativity,
\emph{the phenomena must remain the same in a reference frame moving with a constant speed} \emph{relative to the first one},
and we choose the
system that moves together with the ion. In the second frame, no ionic
movement occurs along the movement's direction. 
In line with the fact that the speed of light is independent of the reference frame,
we assume that the higher interaction speed remains the same in both
systems: it is instant. 
The observers in both reference frames must see that the
system is balanced. The difference is that in the first frame, the
system is \textit{statically} balanced (no change in the gradients, but the
ion is moving), and in the second one, it is \textit{dynamically} balanced (the
gradients change to keep the ion at rest).
\emph{The gradients the
	moving ion experiences are the ones that the standing ion experiences
	at another time (depending on its speed). This way, we can provide
	the needed time course of the process.} 

Compared with the electromagnetic case, we see crucial differences.
First, the mass's propagation speed (forming a new concentration gradient) is millions of times lower than
the charge's. Second, the moving ion simultaneously represents
mass transport and charge transport.
Third, when deriving position derivatives, we conclude from the assumption that
there is no movement (in other words, no explicit dependence on the
time): the effect of the electrical and magnetic driving forces is
equal, whatever time is needed to reach that balanced state. In contrast, in 
electrodiffusion, the velocity changes the concentration gradient, and simultaneously, the
potential gradient.

We assume that equation~(\ref{eq:NernstPlanck}) is valid for a given time
$t$.
\index{Nernst-Planck equation}
At time $t+dt$, in another steady state, the two interactions
manifest at different times: we have
\begin{equation}
	\frac{d}{dx}V_{m}(x+v(x)*dt)=-\frac{RT}{q*F}\frac{1}{C_{k}(x)}\frac{d}{dx}C_{k}(x)\label{eq:Nernst-1V}
\end{equation}
\noindent or, equivalently, it can be expressed as
\begin{equation}
	\frac{d}{dx}C_{k}(x-v(x)*dt)=-\frac{q*F}{RT}C_{k}(x)\frac{d}{dx}V_{m}(x)\label{eq:Nernst-1C}
\end{equation}

\noindent The concentration at position $x$ determines the potential
(apart from an integration constant) at position:
\begin{equation}
	dV_{m}(x)=dx*\frac{d}{dx}V_{m}(x)=-dx\frac{RT}{q*F}\frac{1}{C_{k}(x)}\frac{d}{dx}C_{k}(x)\label{eq:Nernst-2V}
\end{equation}
so (and here the constant disappears), the time derivative is
\begin{equation}
	\frac{d}{dt}V_{m}(x)=v(x)*\frac{d}{dx}V_{m}(x)=-v(x)*\frac{RT}{q*F}\frac{1}{C_{k}(x)}\frac{d}{dx}C_{k}(x)\label{eq:Nernst-dVdt}
\end{equation}
or
\begin{equation}
	\frac{d}{dt}V_m(x)=\frac{D*R}{F}*C(x)*\frac{dC}{dx}*\frac{RT}{q*F}\frac{1}{C(x)}\frac{d}{dx}C_{k}(x)\label{eq:Nernst-dVdt2}
\end{equation}
\noindent Similarly, at time $t-dt$, in another steady state, we
have
\begin{equation}
	dC_{k}(x-v(x)*dt)=dx*\frac{d}{dx}C_{k}(x)=-dx\frac{q*F}{RT}V_{m}(x)\frac{d}{dx}V_{m}(x)\label{eq:Nernst-2C}
\end{equation}
\begin{equation}
	\frac{d}{dt}C_{k}(x)=v(x)*\frac{d}{dx}C_{k}(x)=-v(x)\frac{q*F}{RT}V_{m}(x)\frac{d}{dx}V_{m}(x)\label{eq:Nernst-dCdt}
\end{equation}

We expressed the dependence of gradients on each other using the ion's speed $v$ as an intermediate variable, which can be expressed by the Stokes-Einstein relation as
\begin{equation}
	v(x)= -\frac{D*R}{F}*C(x)*\frac{dC}{dx}
	\label{eq:StokesEinsteinSpeed2}
\end{equation}

\subsection{Fick's Law for electrodiffusion\label{sec:Fick-Electrodiffusion}}

Since the potential is linearly proportional to the concentration (and so are their derivatives), the function is of form $f(x) = - \frac{1}{x}$, where, \textit{as an exception, the square of the first derivative equals the second derivative: $f^{\prime\prime}(x)=(f^{\prime}(x))^2$}.
After simplifying the expression and using that special feature
%
\begin{equation}
	\frac{dV}{dt}=\biggl(\frac{T*R^2}{q*F^2}\biggr)*D*\frac{d^2C}{dx^2}
	\label{eq:FickSecondElectrodiffusion}
\end{equation}

For practical calculations, the voltage's time derivative can be obtained directly from the input current, thereby accounting for the current production mechanism. We must not forget that we started 
from a quasi-equilibrium state; that is, our description is valid only for quasi-static processes.

Given that 
\begin{equation}
	\frac{dV}{dt}= D*\frac{d^2C}{dx^2}\label{Fick'sSecond}
\end{equation}
expresses Fick's Second Law of Diffusion, we can derive 
the ratio between the electrical and thermodynamic temporal gradients.
Using  the values of universal constants
\begin{equation}
	\frac{dV}{dt}=2.23*10^{-6}*D*\frac{d^2C}{dx^2} = \mathbf{2.23*10^{-6}}\frac{dC}{dt}\label{eq:PhysicsGradientRatio}
\end{equation}
We note that non-dedicated experimental results (measuring concentration invasion, see Fig.~2.2 of~\cite{JohnstonWuNeurophysiology:1995}), and voltage invasion, see Fig.~1d of~\cite{BeanActionPotential:2007}, separately), result in an experimental value $~\approx 2*10^{-6}$.

The equation also enables us to understand why the speed
of ion currents (for example, axonal currents) is limited to such low values.
One can assume that the electrical interaction speed is
the respective speed of light, we arrive at the fact
that the speed of material transport (the speed of the mechanical wave) can be up to \SI{120}{[\meter\per\second]}.
(As reviewed in~\cite{MechanicalPropertiesNerves:2025}, the upper limit of sound waves \cite{MechanicalPropertiesNerves:2025}
is about \SI{170}{[\meter\per\second]} in  nerve membranes.
Moreover, the measured value in unmyelinated axons is up to about  \SI{70}{[\meter\per\second]}.)
Since the mechanical wave 
carries electrically charged particles, it induces an electrical charge of opposite sign on the other side
of the axon, which, in addition to the viscosity of the electrolyte, 
brakes the flow (decreases its Stokes-Einstein speed). This way, the particles move under the combined effect of a thermodynamic force and an electrical force. The amount of induced charge depends on the specific capacity 
of the axon (the thickness of the myelin layer on it), so the resultant
force (and, due to this, the Stokes-Einstein speed) of the 
axonal current depends on the thickness of the myelin layer.
As it is known\cite{SolitonMathematics,HEIMBURGReversibleHeatProduction:2021,MechanicalPropertiesNerves:2025}, solitons keep their speed and amplitude while traveling; so the recent biophysical mechanism assuming cooperating ion channels is surely wrong.

For discovering the reciprocal relations in thermodynamics, Lars Onsager was awarded the 1968 Nobel Prize in Chemistry. The presentation speech referred to his result that "Onsager's reciprocal relations represent a further law making a thermodynamic study of irreversible processes possible". In that sense, \textit{our approach provides mathematical equations of the fourth law of thermodynamics}.
The experimental verification~\cite{OnsagerExperimental:1959} of that law mentions "the well-known difficulty of carrying out these experiments": we should have to measure potential changes 
at distances of the fractions of the size of the electrodes, 
with picosecond resolution, while the electrolytic electrodes 
cause nearly $msec$ delays). 
We can overcome that experimental difficulty by using our mathematical relations between electrical and chemical diffusion coefficients.
The significance of our Eq.~(\ref{eq:PhysicsGradientRatio}) is, that one can derive the speed of 
electrodiffusion in electrolytes, which is otherwise not measurable (Classic theory can only mention "hopping in a breeze"~\cite{KochBiophysics:1999}).

\section{Conclusion}\label{sec13}

Ions are 'outlaw' objects of nature in the sense that they are
simultaneously mass and charge. Unlike other objects, they cannot 
be reduced to a single abstraction. In this way, they belong to
two disciplines simultaneously. However, both disciplines
can describe only one of the two abstractions. In other words,
in a disciplinary approach, one of the abstractions always
remains out of the scope of the laws. Thermodynamics discusses 
particles having only mass but no charge, while electricity 
has laws about massless charged particles. Furthermore, 
both disciplines must bridge the continuous and discrete views of science.

The laws, abstracted to the case of one feature
when the other feature is absent, cannot be perfect. Fundamentally,
the direct collisions between discrete particles having masses represent an instant interaction, while their electrostatic repulsion 
represents a remote interaction. The two interaction speeds
differ by several orders of magnitude, and science is not prepared
to handle their combined effect.
Joining the two abstractions that were previously considered independent, 
leads to the establishment of a new scientific discipline.
Examples include thermodynamics, where the discrete and continuous views are combined, and the theory of relativity, which contracts time and space.
Describing ions by thermodynamics needs establishing a connection
between thermodynamics and electricity,
by considering charge and mass simultaneously. Similar to the 
special theory of relativity, the speed can connect them
when deriving their laws of motion.

The cross-disciplinary handling has special importance when
describing living matter. The disciplinary handling failed 
to provide a contradiction-free description of neurons. Our handling,
considering also the "construction" of living matter, succeeded
in providing a unified thermodynamic/electrical model of
biological neurons. 
\backmatter

%
%
%

%
%
%
%

\begin{appendices}

\section{Handling speeds\label{sec:Physics-SpeedHandling}}

\textit{The confusion and question marks in connection with describing life by science mostly arise from the interpretation of the notion 'speed' in physics.}
When discussing the underlying physical laws for biology, we
go back to the fundamental physical concepts instead of taking over the
approximations and abstractions
used in the \textit{classical physics for non-biological matter} and
less complex interactions.
The notions and laws depend
on the circle of phenomena we know and want to describe.
The Newtonian and Einsteinian worlds are
basically distinguished by considering \textit{speed dependence} that actually means \textit{explicit time dependence}.
Interesting consequences are that in the Einsteinian world,
the mass is not constant, time and space are not absolute, and so on.
We can be prepared for some similar counter-intuitive experiences in biology: "we must be prepared to find it working in a manner that cannot be reduced to the \textbf{ordinary} laws of physics"\cite{Schrodinger:1992}.
Here, we scrutinize the fundamental concepts and discover differences
between physics and biology
as consequences of the required different abstractions and approximations.

Physics notoriously suffers from a lack of handling
{different simultaneous interactions}; facing such a case leads to misunderstandings, debates, and causality problems.
Such a famous case is the entanglement speed.
E. Schr\"odinger introduced his famous law of motion in quantum mechanics entirely analogously to how I. Newton introduced his Laws of Motion.
Similar to the Newtonian 'absolute time', the quantum mechanical interaction is supposed to be 'instant'
(this is the price for having 'nice' equations in classical and quantum mechanics),
i.e., its speed is supposed to be infinitely high.
However, at that time, it was already known that the speed of electrical interaction (propagation of electromagnetic waves) is finite.
So if an object has quantum-mechanical interaction (aka entanglement) and electrical interaction simultaneously, the corresponding forces act simultaneously but reach the other object at different times.
The entanglement arrives instantly; the electromagnetic effect arrives at a time we can calculate from the interaction speed and the spatial distance between the objects.
This effect leads to causality problems: the two interactions of photons entangled earlier should be measured at different times, meaning a "spooky remote interaction," as A. Einstein coined, which leads to contradictions such as the Einstein-Podolsky-Rosen paradox.
Actually, the issue stems from the improper handling of mixing interaction speeds:
the Schr\"odinger equation introduces an infinitely large interaction speed,
while the
\gls{EM}
interaction has a finite speed.

During our college studies, we mentioned that light is an electromagnetic
wave with a vast but finite propagation speed. Still, we forgot to
highlight that, at the same time, it is the propagation speed of the
electrical (, and magnetical, and gravitational) interaction force fields
as well. The effect of "Retarded-Time Potential" is also known
in physics and communication engineering. Algorithms "marching-on-in
time" and "Analytical Retarded-Time Potential Expressions" are
derived to handle the problem; for a discussion, see \cite{RetardedTimePotential:2011}.
\index{telegrapher's equations}
Telegrapher's equations (unfortunately, also used to describe biological signal transfer)
explicitly assume a finite propagation speed, millions of times slower
than the (implicitly assumed)
{EM} 
interaction's. The issue is not confined to large
distances: designers of micro-electronic devices also must consider
the effect: they introduced clock time domains and clock distribution
trees; see, for example \cite{WiringDominance:2019, VeghRevisingClassicComputing:2021}.

Science uses 'instant' in the sense that one interaction
is much faster than the process under study; we consider the faster interaction as instant.
\index{instant interaction}
The approach of classical science is based on the oversimplified
approximation that the interaction speed is \emph{always} much higher
than the speed of changes it causes and that the processes can \emph{always}
be described by a single stage. In our approach, for biology, we put together a
\textit{series of stages} to describe the observed complex phenomena, where the stages provide input and output
for each other, involve more than one interaction speed, and use per-stage valid
approximations. We simplify the approximations by omitting the less
significant interactions and introduce ideas for accounting for the different
interaction speeds. This way, we reduce the problem to a case that
science can describe mathematically. \emph{This procedure is fundamentally different
	from applying some mathematical equations derived for an abstracted
	case of science to a complex biological phenomenon without validating
	that we use the appropriate formalism}.

\section{Thermodynamics and statistical physics\label{sec:statistical}}

Thermodynamics is a somewhat misunderstood branch of physics. As its name suggests, it deals with heat, work, and temperature (notice that \textit{charge is not included}), and their relationship to energy, entropy, and the physical properties of matter.
It uses probability-theoretic concepts that may not apply to the system under study.
The connection between the two fields was established by Ludwig Boltzmann, who treated the number of particles in an element of phase space, calculated from a probability distribution, as a real number and then applied the laws governing those particles.

From the point of view of the above dichotomies, it combines the continuous and the discrete views, combines reversible and irreversible processes, but excludes electricity.
The essential points of Boltzmann's general equation are that \textit{the force acting on the particles is instantaneous, the particles only come into contact by direct collisions, the volume is infinite, the number of particles is sufficiently large, and the system is closed}.
Only with these assumptions can we assume that the volume in the phase space does not change. In the case of neutral atoms or molecules in a large, closed volume, the conditions are met, and Boltzmann's equations describe the observed behavior of the systems.
Another approach can be derived (the so-called Vlasov equations) when there is a long-range interaction between particles (for example, the Coulomb interaction between ions of the same charge), which excludes direct particle collisions.
The latter describes, for example, plasmas in which atoms are fully ionized and ions move through a cloud of free electrons. This model also partially describes electrons moving in the periodic force field of the ion lattice in a solid, but it is not considered a thermodynamic system in the former sense; that concept is reserved for particles that interact only by direct collisions. 

When applied to ions and especially to electrolytes, the conditions for the applicability of the Boltzmann equations are largely not met. Due to the low temperature compared to plasmas, the vast majority of molecules do not dissociate, so they interact only through direct collisions. However, a minimal number of ions, due to their identical charge, avoids direct collisions through long-range interactions.
They do, however, collide with neutral molecules. Ions play leading roles in biological processes and can therefore interact through both direct collisions and long-range forces. For this reason, \textit{the Boltzmann equations are certainly not applicable to the description of the processes taking place in electrolytes}, which directly explains why thermodynamics cannot describe processes of life based on electrolytes.

Further limitations are that, in biological systems, statistical concepts must be applied to a very small number of particles in a closed space of finite volume,
where coercive forces also occur, and processes beyond the experimenter's control make it doubtful whether we can interpret a closed system.
The main problem, however, is that ion charge and mass are inseparable. 
Furthermore, the speed of the electrostatic interaction of the particles is instantaneous, in accordance with Boltzmann's original assumption. 
In contrast, the speed of the diffusional interaction, which is calculated from the change in the distribution of the particles, is orders of magnitude slower. Of course, the particle's motion simultaneously changes the strengths of both the electrical and thermodynamic interactions. The thermodynamic effect occurs only when the particle enters the region where the concentration changes, whereas the force field created by the other particles changes immediately.

Classical physics can only handle instantaneous interactions; in the Newtonian worldview, all interactions co-occur. Overall, the traditional approach to thermodynamics cannot be applied to biological systems. In essence, Erwin Schr\"odinger stated that \textit{ordinary} scientific laws cannot describe living matter. At the same time, he expressed his conviction that no new force or unknown interaction is emerging (the 'whatnot' as he coined likely includes the magic 'protein mechanisms' that move ions against the electrical and thermodynamic forces); only a previously unknown regularity concept must be found, then the non-ordinary laws organically integrate into the fabric of science, together with the already known laws.

To derive laws of motion of ions in electrolytes, one must use non-ordinary (aka cross-disciplinary) methods after scrutinizing which approximation can be used for living matter. Below, we derive an approach that enables us to handle interaction speeds that differ by orders of magnitude. Science has created abstract concepts such as space and time, mass and charge. Modern physics was born when experiments began to contradict the fact that nature could be described in such simple terms. The recognition of the non-continuity of energy led to the creation of quantum mechanics, and the recognition of the non-independence of space and time led to the creation of the theory of relativity, with far-reaching consequences.
We understood that the Newtonian approach has limitations, and that interactions at finite speeds can only be described by a different kind of mathematics, namely the concept of spacetime. Moreover, mass is also related to space and time; taking this into account, we can describe nature in terms of curved spacetime. According to classical physics, we can describe phenomena with sufficient accuracy using only the aforementioned concepts (the zeroth derivatives). According to modern physics, for a more accurate description, we must take into account the first derivative with respect to time. For a general description, we must consider the second derivative as well.

Regarding ions, we can handle their charge and mass separately; we know the relevant laws. However,
in the case of ions, we must connect charge and mass, just as in the case of the theory of relativity, space and time. Interestingly,
here too, the velocity creates the connection between the two abstract features. Furthermore, we must introduce a relation for the currents carried by ions, similar to that introduced in statistical mechanics for the correspondence between the particle and the continuum views. However, in our case, the correspondence is established on geometrical grounds.

\section{Biology vs physics\label{sec:Physics-BiologyVsPhysics}}

A common fallacy in biology is that 
{physics} cannot underpin the operation of living matter, citing E.~Schr\"odinger. However, the claim falsifies his opinion by omitting the most essential word, 'ordinary'. 
Schr\"odinger wanted to emphasize the opposite: there is no new force (no unknown new interaction, as biophysics attempts to introduce "protein mechanisms"), only that studying living matter requires different testing methods (and we add: different uses of physics' concepts) in the physical laboratory. He suggested answering the question
"Is life based on the laws of physics?" affirmatively, but expected to discover the appropriate forms of physical laws describing the 'non-ordinary' (in our reading: non-disciplinary) behavior of living matter. No doubt, the basic concepts and terms must be interpreted precisely for living matter,
much beyond the level we used to at the college level.
However, after that reinterpretation, we can interpret features of living matter, although we need a more careful, cross-disciplinary analysis to do so. 
We must use appropriate abstractions and approximations for the phenomena,
depending on the level required for the given cooperation of objects and interactions.
Furthermore, we discuss some of the relevant terms and notions of physics,
differentiating which approximation is appropriate only for physics (mainly electricity),
and,  instead, which approximation should be used for biology.
As we discuss, \textit{biophysics translated the corresponding major terminus technicus words
	from the theory and practice of physics' major disciplines, mainly from electricity, which were
	worked out for homogeneous, isotropic, structureless metals,
	and for strictly pair-wise interactions with a single (actually, 'instant') interaction speed; to the structured,
	\index{interaction!attributes}
	non-homogeneous, non-isotropic, material mixtures and for multiple interaction speeds}.
Those notions rarely retain their original meanings, and how much they do depends on the actual conditions. 
The precise meaning needs a case-by-case analysis.

The physical models consider infinitely large volumes,
surfaces, distances; furthermore, and most importantly, instant interactions. Is the cell large enough to
consider it infinitely large (at least on the scale using ions' size); that is, to apply laws of science
derived for the abstraction 'infinitely large'? 
When working with charge, we know that charge is quantized, while the macroscopic quantities voltage and current are 
continuous (derivable). Do cells contain a sufficient 
number of charge carriers to apply macroscopic notions?
Do the thousands of times smaller ion channels transfer enough charge
to speak about a well-defined current?
When a couple of ions are transferred through an ion channel,
do they significantly change the potential that accelerates them?

Science could serve as a firm base for all its disciplines.
As we discuss, \textit{its disciplines use abstractions based on limited-validity approximations} based on the same first principles.
However, \textit{the approximations differ between biology and physics}.
In physics, some processes we observe are fast enough that we can treat them as essentially state transitions.	In some cases, the approach can
be --more or less-- successful. For the slower, well-observable processes,
we have the 
laws of motion
that describe how processes occur under the influence of a driving force.
We also found that nature is not necessarily linear (in the sense that it depends only on the mentioned quantities, not on their derivatives), which we can describe with "nice" mathematical formulas.
A century ago, A.~Einstein discovered that the approximations I.~Newton introduced two centuries earlier are not sufficiently accurate for describing the movement of bodies at high speeds. In other words, a new paradigm, the constancy of the speed of light, must have been introduced
that caused a revolution in physics and led to the birth of "modern physics". 

\textit{Life, including the brain's operation, is dynamic.} 
As Schr\"odinger formulated, the "construction of living matter" differs
from the one science used to test in its labs.
The scientific abstraction based on "states" (i.e., on instant changes)
fails for the case of biology, where "processes" happen (i.e., the changes
are obviously much slower).
The commonly used measuring methods, such as {clamping, patching, and freezing},  reduce the life to states. 
The related theories describe states with perturbation~\cite {PerturbationNeuralComputation:2002}. 
On the one hand, this technology fixes the cell in a well-defined, static state, enabling us to observe a static anatomic picture of the cell. On the other hand,
it eliminates the dynamic processes from the theory, i.e., \textit{hides forever the essence of the life that the cell exists in a continuous change governed by laws of motion}.
Those methods stop the processes under study, thereby depriving them of their dynamicity and preventing measurement.
It was forgotten that using feedback for stabilizing an autonomously
working electrical system means introducing foreign currents,
and this way falsifying its operation. 

Furthermore, it is hazardous to introduce technically (and incorrectly) derived
and misinterpreted macroscopic features and interpret them as fundamental
electrical notions. In general, instead of understanding and developing the proper scientific basis for the operation, they say that science cannot describe it. The idea of {conductance}
has been introduced
to neurophysiology almost a century ago. It was taken from physics, where the
notion was derived for metals (conductors, instead of electrolytes). Since then, its original interpretation
has been forgotten, and today (in contrast with physics), it has become
a primary entity for describing electrical characteristics of biological
cells. We explain how the right physics background enables us to
discover wrong physical models and 
misinterpreted notions of physics in neurophysiology.
Furthermore, the proper interpretation opens the way to the correct interpretation
of neuronal information. We set up an abstract electrical/thermodynamic model of neuronal
operation.

We derived the needed 'non-ordinary laws'~\cite{VeghNon-ordinaryLaws:2025}, which are derived by using the same first principles as the 'ordinary laws', but are abstracted for the
approximations valid for living matter.
As we discuss, those 'ordinary' laws were derived for strictly pair-wise interactions at very high speeds.
In biology, we can observe interactions at a million times lower speed, in inhomogeneous, non-isotropic, structured material.
\textit{Biology does not have the conditions for which physics derived its ordinary laws.}
By using the appropriate approximations for the 
biological cases, we can derive the required 'non-ordinary' laws of physics,
which laws are more complex to derive and use; furthermore, we must use several
stages (with the approximations changing from stage to stage) instead of 
one single stage, as in the case of the 'ordinary' laws. 
However, \textit{all laws follow the same principles}.

Biology, and predominantly neuronal operation, produces examples where using wrong omissions
in complex processes results in absolutely wrong results. In those cases,
some initial resemblance between our theoretical predictions and our phenomena exists. However, the success
in simple cases provides no guarantee that the model was appropriate: "the success of the equations is no evidence in
favour of the mechanism"~\cite{HodgkinHuxley:1952}. 
\textit{Finally,
	all laws are approximations, and the accuracy of verifying their predictions is limited.}
Several theories can describe the same phenomenon with the required accuracy.
We also show in the section about the 
{finite interaction speeds}
that the most well-known laws (from Newton, Coulomb, Kirchoff, etc.) are
also approximations. They have their range of validity, although it is often
forgotten.

One such neuralgic point of omissions and approximations is the vastly different
{interaction speeds}. Furthermore, where the speed is considered at all, \textit{the same speed is assumed for all interactions}.
The laws are abstract also in the sense that, say, the objects in the laws of physics have either mass or
electrical charge, but not both. It is the researcher's task to decide
which combination of laws
can be applied to the given condition. For example, one can assume in most cases
that the speeds sum up linearly, except at very high speeds.
Biology provides excellent case studies where different interactions
shape the phenomenon, and special care must be exercised.
We give a short review of 
{history and kinds of interaction speeds}.

Another point is that science started with the assumption that non-living matter is continuous, although it was soon discovered that it consists of the smallest particles.
When we reached that size, we experienced that different subsets
of science laws describe that matter and the atoms they contain.
It is one of the most challenging tasks to establish relations between
those subsets. Again, we used abstractions that the 
continuous matter is infinitely large and that the isolated
atoms are infinitely far from each other and from the external world. We also experienced the semi-infinite cases, and studied
the behavior of surfaces and interfaces, which, again, is different from both that of the atoms and their large masses. Given that biological objects span the microscopic to macroscopic size range and are surrounded by surfaces, we must be prepared for the fact that no simple rules describe their behavior.

Neuronal operation is at the boundary, where sometimes, in the same phenomenon,
one interaction can be interpreted at the macroscopic level, another
must already be interpreted at the microscopic level.
Furthermore, a series of stages (instead of a single state) and
processes (instead of stages) describes the subject under study.
Given the vital role of 
charge and current in neuronal operation, we provide their precise interpretations.
Furthermore, we must consider that the processes happen in a finite volume,
"within the spatial boundary".

\section{Speculations\label{sec:Physics-Speculations}}
By using the value of the force acting on a unit charge
in the field across the membrane is
\begin{equation}
	F_{Na^+}= 10^7 *1.60217663* 10^{-19} [V] [C] = 1.6*10^{-12}\ [N] \label{eq:UnitForce}
\end{equation}
we can estimate how the pressure of the neural cell increases due to the rush-in changes at the beginning of the
\gls{AP}. 
As evidence shows, the local potential at the internal surface of the membrane is in the range of \SI{100}{\milli\volt} in the resting state and increases by $\Delta U=$ \SI{100}{\milli\volt}  in the transition state. This increase means a change in the force acting on an ion (see Eq.~(\ref{eq:UnitForce})) by  $1.6*10^{-12}\ [N]$.
When we assume $10^7$ rush-in ions and unchanged cell size, the total force acting on the membrane increases by $1.6*10^{-5}\ [N]$.
This change in force means on the neuron's $8*10^{-9}\ [m^2]$ surface (see~\cite{JohnstonWuNeurophysiology:1995}, page~12) a pressure change 
\begin{equation}
	\Delta P = \frac{1.6*10^{-5}\ [N]}{8*10^{-9}\ [m^2]}
	= 2*10^{3}\quad \biggl[\frac{N}{m^2}\biggr]
	\label{eq:CellPressureChange}
\end{equation}
\noindent Again, the pressure can be calculated 
using electricity.

This pressure on a $10\times10~\mu m$ tooltip is converted to a force
$2*10^{3}*100*10^{-12} = 200~pN$ force. The measured force value~\cite{MechanicalPropertiesNerves:2025} is about $600~pN$,
so our estimation is in the correct range. 
The measured pressure value is~\cite{PressureChangeActionPotentialMeasured:1980} $5\ \frac{dyn}{cm^2} = 0.5\ \frac{N}{m^2}$, which is an average value
for some period. If we assume a $1~Hz$ frequency, and a $1~ms$
period for the duration of the \gls{AP} peak, it means $1*10^{3}\, \bigl[\frac{N}{m^2}\bigr]$.
So, the measured mechanical change values~\cite{NeuronalDeformation:2020} do not contradict our hypothesis that the increased pressure increases the neuron's size via its elasticity.

A way to estimate the ions' rush-in speed is 
that one assumes that the Stokes-Einstein speed describes
the ions' speed (Fig.~3 in~\cite{HodgkinHuxley:1952}, quantitatively underpins the hypothesis, see~\cite{VeghStokesEinstein:2025}) both in axons and ion channels. If so, 
the ratio of the potential gradients equals the ratio of the 
speeds of ions. One can estimate that the $100~mV$ \gls{AP} on the 
$50~\mu m$ \gls{AIS} generates a $2*10^4~\frac{V}{m}$ electrical gradient and 
\cite{HodgkinHuxley:1952} measured $20~\frac{m}{s}$ for the speed of the \gls{AP}. By assuming that the electrical field across the membrane is $10^7~\frac{V}{m}$, one can estimate the speed at the exit of the ion channels as $10^4~\frac{m}{s}$.

We can also estimate the work done when the rush-in charge
extends the neuron, as the $\Delta P\times\Delta V$. We consider that $\Delta P$ remains constant, and we calculate the change of the volume as neuron's surface $10^{-8}\ m^2$ multiplied by 
the change of the radius $\Delta r=10^{-9}\ [m]$ multiplied by the pressure $\Delta P$, that gives $E_{mech}=2*10^{-14}\ [J]$.
If one assumes that the pressure is of entirely mechanical origin~\cite{NeuronalDeformation:2020}, one must assume that the 
pressure is $10^6~\frac{N}{m^2}$, which is well above the measured value.
We can also estimate the electrical energy $E_{electr}=\frac{1}{2} C*(\Delta V)^2 = 1.4*10^{-12}\ [J]$, see also~\cite{EnergyNeuralCommunication:2021}.
(This conclusion is in line with that "the magnitude of the capacitive energy is much smaller than the heat exchange that is observed", as detailed in~\cite{MechanicalPropertiesNerves:2025}.)
To estimate the time required to change the pressure, one can assume that the collision force of a $Na^+$ ion provides the pressure (the shock wave), and this can be calculated as $F = m\frac{\Delta v}{\Delta t}$. Given that Eq.~(\ref{eq:UnitForce})  provides the force, and one must assume that the ion loses its 
$10^4~to~10^5\frac{m}{s}$ speed, one concludes that it generates a shock wave in $\Delta t = 10^{-8}~to~10^{-7}~seconds$.

The consequences of the changes in the electrical charge are large enough to explain why 
pressure wave and other mechanical changes~\cite{MechanicalWaves:2015} also start at the beginning of the
\gls{AP},
and other (such as optical, density) changes 
are accompanied by it. Nature invests energy also 
in the conventional way, a term $V\Delta P$, into the thermodynamics of neural operation; not only in the form of storing energy in the 
changed electrical field. Thermodynamics and electricity, not to mention elasticity, must not be 
separated when discussing the neuronal energy business. 

In line with the experimentally measured energy consumption,
the \gls{AP} cycle consumes energy in the range of $10^{-7}\ [J]$. 
The five orders of magnitude between the total and the electrical energy
seems to underpin that in addition to the $10^{-3}$ concentration of ions,
only a small fraction of the ions (the ones in the layers near the membrane) in the volume participate in the electric activity of issuing an \gls{AP}, while the entire volume in the mechanical activity.
That is, practically all the energy is invested as elastic energy, but the elasticity modulus of the membrane is exceptionally high.
The primary reason for neuronal operation is the appearance of new ions in the vicinity of the membrane, but this creates a secondary, vast mechanical shock wave. By considering the electrical repulsion force between particles in a slow current, which is entirely omitted in the classic theory, we open the way for explaining the observed thermodynamic and mechanical changes. Instead of alternative disciplinary theories, cooperation between the classic disciplines is needed.

For thermodynamic distributions, one can interpret the "temperature of individual particles" with mean kinetic energy $E$ as $T=\frac{2*E}{3*k_B}$ (where $k_B$ is the Boltzmann constant and $T$ is the thermodynamic temperature of the
bulk quantity of the substance); that is, the temperature is directly proportional to the average kinetic energy.
The energy arrives at the neuron in the form of $10^7$ $Na^+$ ions,
so the average energy a single ion consumes is $2.5*10^{-14}~[J]$. One can assume that most (99\%) of that energy is consumed for moving
the ion against the Stokes-Einstein force in the viscous medium, 
so after exiting the membrane, the speed of the ions is $\sqrt{\frac{2*2.5*10{-16}}{3.82*10^{-26}}}\approx 10^5~[m/s]$, 
which is well above the $\approx~500~m/s$ average speed.
The ions on the two sides of the membrane are in the state with temperature $T$ before and after the \gls{AP}. When the $Na^+$ ions rush into the intracellular space, they gain energy through electrostatic acceleration by the membrane's electrical potential, so the same ions appear on the intracellular side as having more energy, i.e., slightly higher temperature (the temperature is more than two orders of magnitude higher, but the proportion of those "hot" ions is about six orders of magnitude lower, so one can expect temperature change up to the range of fragments of up to millikelvin, depending on the measuring conditions).
In the second phase of the \gls{AP}, those more energetic ions that provide excess voltage (and so: excess local potential) above the resting potential leave the membrane's proximity through the \gls{AIS},
so the temperature decreases in the second phase to its original value 
It is another effect that can contribute to the general heat production and adsorption discussed above.
In the original publication on solitons (essentially pressure waves)~\cite{SolitonPropagation:2005}, a wide range of measured and theoretical data is discussed. They are in line with our conclusions, among others, $80\ \mu K$ measured temperature change.
Furthermore, the electrical energy is more than an order of magnitude smaller than the elastic energy (called soliton energy there).

\end{appendices}



\begin{thebibliography}{29}
	\ifx \bisbn   \undefined \def \bisbn  #1{ISBN #1}\fi
	\ifx \binits  \undefined \def \binits#1{#1}\fi
	\ifx \bauthor  \undefined \def \bauthor#1{#1}\fi
	\ifx \batitle  \undefined \def \batitle#1{#1}\fi
	\ifx \bjtitle  \undefined \def \bjtitle#1{#1}\fi
	\ifx \bvolume  \undefined \def \bvolume#1{\textbf{#1}}\fi
	\ifx \byear  \undefined \def \byear#1{#1}\fi
	\ifx \bissue  \undefined \def \bissue#1{#1}\fi
	\ifx \bfpage  \undefined \def \bfpage#1{#1}\fi
	\ifx \blpage  \undefined \def \blpage #1{#1}\fi
	\ifx \burl  \undefined \def \burl#1{\textsf{#1}}\fi
	\ifx \doiurl  \undefined \def \doiurl#1{\url{https://doi.org/#1}}\fi
	\ifx \betal  \undefined \def \betal{\textit{et al.}}\fi
	\ifx \binstitute  \undefined \def \binstitute#1{#1}\fi
	\ifx \binstitutionaled  \undefined \def \binstitutionaled#1{#1}\fi
	\ifx \bctitle  \undefined \def \bctitle#1{#1}\fi
	\ifx \beditor  \undefined \def \beditor#1{#1}\fi
	\ifx \bpublisher  \undefined \def \bpublisher#1{#1}\fi
	\ifx \bbtitle  \undefined \def \bbtitle#1{#1}\fi
	\ifx \bedition  \undefined \def \bedition#1{#1}\fi
	\ifx \bseriesno  \undefined \def \bseriesno#1{#1}\fi
	\ifx \blocation  \undefined \def \blocation#1{#1}\fi
	\ifx \bsertitle  \undefined \def \bsertitle#1{#1}\fi
	\ifx \bsnm \undefined \def \bsnm#1{#1}\fi
	\ifx \bsuffix \undefined \def \bsuffix#1{#1}\fi
	\ifx \bparticle \undefined \def \bparticle#1{#1}\fi
	\ifx \barticle \undefined \def \barticle#1{#1}\fi
	\bibcommenthead
	\ifx \bconfdate \undefined \def \bconfdate #1{#1}\fi
	\ifx \botherref \undefined \def \botherref #1{#1}\fi
	\ifx \url \undefined \def \url#1{\textsf{#1}}\fi
	\ifx \bchapter \undefined \def \bchapter#1{#1}\fi
	\ifx \bbook \undefined \def \bbook#1{#1}\fi
	\ifx \bcomment \undefined \def \bcomment#1{#1}\fi
	\ifx \oauthor \undefined \def \oauthor#1{#1}\fi
	\ifx \citeauthoryear \undefined \def \citeauthoryear#1{#1}\fi
	\ifx \endbibitem  \undefined \def \endbibitem {}\fi
	\ifx \bconflocation  \undefined \def \bconflocation#1{#1}\fi
	\ifx \arxivurl  \undefined \def \arxivurl#1{\textsf{#1}}\fi
	\csname PreBibitemsHook\endcsname
	
	\bibitem[\protect\citeauthoryear{Schr\"odinger}{1992}]{Schrodinger:1992}
	\begin{bbook}
		\bauthor{\bsnm{Schr\"odinger}, \binits{E.}}:
		\bbtitle{Is life based on the laws of physics?},
		pp. \bfpage{76}--\blpage{85}.
		\bpublisher{Cambridge University Press},
		\blocation{Canto}
		(\byear{1992})
	\end{bbook}
	\endbibitem
	
	\bibitem[\protect\citeauthoryear{Feynman}{}]{FeynmanThinking:1980}
	\begin{botherref}
		\oauthor{\bsnm{Feynman}, \binits{R.P.}}:
		Principles of scientific thinking.
		\url{https://blog.hptbydts.com/richard-feynmans-principles-of-scientific-thinking}
	\end{botherref}
	\endbibitem
	
	\bibitem[\protect\citeauthoryear{Drukarch
		et~al.}{2022}]{ThermodynamicAPDrukarch:2022}
	\begin{barticle}
		\bauthor{\bsnm{Drukarch}, \binits{B.}},
		\bauthor{\bsnm{Wilhelmus}, \binits{M.M.M.}},
		\bauthor{\bsnm{Shrivastava}, \binits{S.}}:
		\batitle{The thermodynamic theory of action potential propagation: a sound
			basis for unification of the physics of nerve impulses}.
		\bjtitle{Reviews in the Neurosciences}
		\bvolume{33}(\bissue{3}),
		\bfpage{285}--\blpage{302}
		(\byear{2022})
		\doiurl{10.1515/revneuro-2021-0094}
	\end{barticle}
	\endbibitem
	
	\bibitem[\protect\citeauthoryear{Hodgkin and Huxley}{1952}]{HodgkinHuxley:1952}
	\begin{barticle}
		\bauthor{\bsnm{Hodgkin}, \binits{A.L.}},
		\bauthor{\bsnm{Huxley}, \binits{A.F.}}:
		\batitle{A quantitative description of membrane current and its application to
			conduction and excitation in nerve}.
		\bjtitle{J. Physiol.}
		\bvolume{117},
		\bfpage{500}--\blpage{544}
		(\byear{1952})
	\end{barticle}
	\endbibitem
	
	\bibitem[\protect\citeauthoryear{Pence}{2017}]{HH_Potential_Controversies_2017}
	\begin{barticle}
		\bauthor{\bsnm{Pence}, \binits{D.E.}}:
		\batitle{{Potential Controversies: Causation and the Hodgkin and Huxley
				Equations}}.
		\bjtitle{Philosophy of Science}
		\bvolume{84}(\bissue{5}),
		\bfpage{1177}--\blpage{1188}
		(\byear{2017})
		\doiurl{10.1086/694040}
	\end{barticle}
	\endbibitem
	
	\bibitem[\protect\citeauthoryear{Hodgkin}{1964}]{HodgkinConduction:1964}
	\begin{bbook}
		\bauthor{\bsnm{Hodgkin}, \binits{A.L.}}:
		\bbtitle{The Conduction of the Nervous Impulse}.
		\bpublisher{Liverpool University Press},
		\blocation{Liverpool, UK, 1964}
		(\byear{1964})
	\end{bbook}
	\endbibitem
	
	\bibitem[\protect\citeauthoryear{Abbott
		et~al.}{1958}]{HeatProductionNeuron:1958}
	\begin{barticle}
		\bauthor{\bsnm{Abbott}, \binits{B.C.}},
		\bauthor{\bsnm{Hill}, \binits{A.V.}},
		\bauthor{\bsnm{Howarth}, \binits{J.V.}}:
		\batitle{The positive and negative heat production associated with a nerve
			impulse.}
		\bjtitle{Proc. R. Soc. London. B.}
		\bvolume{148},
		\bfpage{149}--\blpage{187}
		(\byear{1958})
	\end{barticle}
	\endbibitem
	
	\bibitem[\protect\citeauthoryear{Levy and
		Calvert}{2021}]{EnergyNeuralCommunication:2021}
	\begin{barticle}
		\bauthor{\bsnm{Levy}, \binits{W.B.}},
		\bauthor{\bsnm{Calvert}, \binits{V.G.}}:
		\batitle{{Communication consumes 35 times more energy than computation in the
				human cortex, but both costs are needed to predict synapse number}}.
		\bjtitle{Proceedings of the National Academy of Sciences}
		\bvolume{118}(\bissue{18}),
		\bfpage{2008173118}
		(\byear{2021})
		\doiurl{10.1073/pnas.2008173118}
	\end{barticle}
	\endbibitem
	
	\bibitem[\protect\citeauthoryear{El~Hady and
		Machta}{2019}]{MechanicalWaves:2015}
	\begin{barticle}
		\bauthor{\bsnm{El~Hady}, \binits{A.}},
		\bauthor{\bsnm{Machta}, \binits{B.B.}}:
		\batitle{{Mechanical surface waves accompany action potential propagation}}.
		\bjtitle{Nature Communications}
		\bvolume{6},
		\bfpage{6697}
		(\byear{2019})
		\doiurl{10.1038/ncomms7697}
	\end{barticle}
	\endbibitem
	
	\bibitem[\protect\citeauthoryear{Miller}{1959}]{OnsagerExperimental:1959}
	\begin{botherref}
		\oauthor{\bsnm{Miller}, \binits{D.G.}}:
		{THERMODINAMICS OF IRREVERSIBLE PROCESSES THE EXPERIMENTAL VERIFICATION OF THE
			ONSAGER RECIPROCAL RELATIONS }.
		Technical Report Contract No. W-7405-eng-48,
		UNIVERSITY OF CALIFORNIA, Lawrence Radiation Laboratory, Livermore, California
		(1959).
		\doiurl{10.1021/cr60203a003}
	\end{botherref}
	\endbibitem
	
	\bibitem[\protect\citeauthoryear{Heimburg}{2007}]{ThermalBiophysics:2007}
	\begin{bbook}
		\bauthor{\bsnm{Heimburg}, \binits{T.}}:
		\bbtitle{Thermal Biophysics of Membranes},
		\bedition{1}st edn.
		\bpublisher{2007 WILEY-VCH Verlag GmbH \& Co. KGaA},
		\blocation{Weinheim}
		(\byear{2007})
	\end{bbook}
	\endbibitem
	
	\bibitem[\protect\citeauthoryear{V{\'e}gh}{2025}]{VeghNon-ordinaryLaws:2025}
	\begin{botherref}
		\oauthor{\bsnm{V{\'e}gh}, \binits{J.}}:
		{The 'non-ordinary laws' of physics describing life}.
		{arXiv}, in review in Foundations of Physics
		\textbf{1}
		(2025)
	\end{botherref}
	\endbibitem
	
	\bibitem[\protect\citeauthoryear{Koch}{1999}]{KochBiophysics:1999}
	\begin{bbook}
		\bauthor{\bsnm{Koch}, \binits{C.}}:
		\bbtitle{Biophysics of Computation}.
		\bpublisher{Oxford University Press},
		\blocation{New York, Oxford}
		(\byear{1999})
	\end{bbook}
	\endbibitem
	
	\bibitem[\protect\citeauthoryear{V{\'e}gh}{2025}]{VeghMembranePotential:2025}
	\begin{botherref}
		\oauthor{\bsnm{V{\'e}gh}, \binits{J.}}:
		The unified non-disciplinary model of the operation of neurons.
		arXiv, in review in Journal of Biological Physics,
		1--76
		(2025)
		\doiurl{10.48550/arXiv.2507.11448}
	\end{botherref}
	\endbibitem
	
	\bibitem[\protect\citeauthoryear{Kandel
		et~al.}{2013}]{PrinciplesNeuralScience:2013}
	\begin{bbook}
		\bauthor{\bsnm{Kandel}, \binits{E.R.}},
		\bauthor{\bsnm{Schwartz}, \binits{J.H.}},
		\bauthor{\bsnm{Jessell}, \binits{T.M.}},
		\bauthor{\bsnm{Siegelbaum}, \binits{S.A.}},
		\bauthor{\bsnm{Hudspeth}, \binits{A.J.}}:
		\bbtitle{Principles of Neural Science},
		\bedition{5}th edn.
		\bpublisher{The McGraw-Hill Medical},
		\blocation{New York Chicago etc.}
		(\byear{2013})
	\end{bbook}
	\endbibitem
	
	\bibitem[\protect\citeauthoryear{Heimburg and
		Jackson}{2005}]{SolitonPropagation:2005}
	\begin{botherref}
		\oauthor{\bsnm{Heimburg}, \binits{T.}},
		\oauthor{\bsnm{Jackson}, \binits{A.D.}}:
		On soliton propagation in biomembranes and nerves.
		Proceedings of the National Academy of Sciences,
		9790--9795
		(2005)
		\doiurl{10.1073/pnas.0503823102}
	\end{botherref}
	\endbibitem
	
	\bibitem[\protect\citeauthoryear{Heimburg}{2025}]{MechanicalPropertiesNerves:2025}
	\begin{barticle}
		\bauthor{\bsnm{Heimburg}, \binits{T.}}:
		\batitle{The mechanical properties of nerves, the size of the action potential,
			and consequences for the brain}.
		\bjtitle{Chemistry and Physics of Lipids}
		\bvolume{267},
		\bfpage{105461}
		(\byear{2025})
		\doiurl{10.1016/j.chemphyslip.2024.105461}
	\end{barticle}
	\endbibitem
	
	\bibitem[\protect\citeauthoryear{Forcella
		et~al.}{2014}]{ViscousChargedFluids:2014}
	\begin{barticle}
		\bauthor{\bsnm{Forcella}, \binits{D.}},
		\bauthor{\bsnm{Zaanen}, \binits{J.}},
		\bauthor{\bsnm{Valentinis}, \binits{D.}},
		\bauthor{\bsnm{Marel}, \binits{D.}}:
		\batitle{Electromagnetic properties of viscous charged fluids}.
		\bjtitle{Phys. Rev. B}
		\bvolume{90},
		\bfpage{035143}
		(\byear{2014})
		\doiurl{10.1103/PhysRevB.90.035143}
	\end{barticle}
	\endbibitem
	
	\bibitem[\protect\citeauthoryear{Johnston and
		Wu}{1995}]{JohnstonWuNeurophysiology:1995}
	\begin{bbook}
		\bauthor{\bsnm{Johnston}, \binits{D.}},
		\bauthor{\bsnm{Wu}, \binits{S.M.-S.}}:
		\bbtitle{Foundations of Cellular Neurophysiology}.
		\bpublisher{Massachusetts Institute of Technology},
		\blocation{Cambridge, Massachusetts and London, England}
		(\byear{1995})
	\end{bbook}
	\endbibitem
	
	\bibitem[\protect\citeauthoryear{Bean}{2007}]{BeanActionPotential:2007}
	\begin{botherref}
		\oauthor{\bsnm{Bean}, \binits{B.P.}}:
		The action potential in mammalian central neurons.
		Nature Reviews Neuroscience
		\textbf{8}
		(2007)
		\doiurl{10.1038/nrn2148}
	\end{botherref}
	\endbibitem
	
	\bibitem[\protect\citeauthoryear{Manukure and
		Booker}{2021}]{SolitonMathematics}
	\begin{barticle}
		\bauthor{\bsnm{Manukure}, \binits{S.}},
		\bauthor{\bsnm{Booker}, \binits{T.}}:
		\batitle{A short overview of solitons and applications}.
		\bjtitle{Partial Differential Equations in Applied Mathematics}
		\bvolume{4},
		\bfpage{100140}
		(\byear{2021})
		\doiurl{10.1016/j.padiff.2021.100140}
	\end{barticle}
	\endbibitem
	
	\bibitem[\protect\citeauthoryear{Heimburg}{2021}]{HEIMBURGReversibleHeatProduction:2021}
	\begin{barticle}
		\bauthor{\bsnm{Heimburg}, \binits{T.}}:
		\batitle{The important consequences of the reversible heat production in nerves
			and the adiabaticity of the action potential}.
		\bjtitle{Progress in Biophysics and Molecular Biology}
		\bvolume{162},
		\bfpage{26}--\blpage{40}
		(\byear{2021})
		\doiurl{10.1016/j.pbiomolbio.2020.07.007} .
		\bcomment{On the Physics of Excitable Media. Waves in Soft and Living Matter,
			their Transmission at the Synapse and their Cooperation in the Brain}
	\end{barticle}
	\endbibitem
	
	\bibitem[\protect\citeauthoryear{Ulku and
		Ergin}{2011}]{RetardedTimePotential:2011}
	\begin{barticle}
		\bauthor{\bsnm{Ulku}, \binits{H.A.}},
		\bauthor{\bsnm{Ergin}, \binits{A.A.}}:
		\batitle{Application of analytical retarded-time potential expressions to the
			solution of time domain integral equations}.
		\bjtitle{IEEE Transactions on Antennas and Propagation}
		\bvolume{59}(\bissue{11}),
		\bfpage{4123}--\blpage{4131}
		(\byear{2011})
		\doiurl{10.1109/TAP.2011.2164180}
	\end{barticle}
	\endbibitem
	
	\bibitem[\protect\citeauthoryear{Ho and Horowitz}{2007}]{WiringDominance:2019}
	\begin{botherref}
		\oauthor{\bsnm{Ho}, \binits{R.}},
		\oauthor{\bsnm{Horowitz}, \binits{M.}}:
		{More about wires and wire models}.
		\url{https://web.stanford.edu/class/archive/ee/ee371/ee371.1066/lectures/lect_09_1up.pdf}
		(accessed on 29 April 2025)
		(2007)
	\end{botherref}
	\endbibitem
	
	\bibitem[\protect\citeauthoryear{V\'egh}{2021}]{VeghRevisingClassicComputing:2021}
	\begin{botherref}
		\oauthor{\bsnm{V\'egh}, \binits{J.}}:
		Revising the classic computing paradigm and its technological implementations.
		Informatics
		\textbf{8}(4)
		(2021)
		\doiurl{10.3390/informatics8040071}
	\end{botherref}
	\endbibitem
	
	\bibitem[\protect\citeauthoryear{Maass
		et~al.}{2002}]{PerturbationNeuralComputation:2002}
	\begin{barticle}
		\bauthor{\bsnm{Maass}, \binits{W.}},
		\bauthor{\bsnm{Natschl\"ager}, \binits{T.}},
		\bauthor{\bsnm{Markram}, \binits{H.}}:
		\batitle{Real-time computing without stable states: A new framework for neural
			computation based on perturbations}.
		\bjtitle{Neural Computation}
		\bvolume{14}(\bissue{11}),
		\bfpage{2531}--\blpage{2560}
		(\byear{2002})
		\doiurl{10.1162/089976602760407955}
	\end{barticle}
	\endbibitem
	
	\bibitem[\protect\citeauthoryear{{Iwasa, K and Tasaki, I. and Gibbons,
			RC.}}{1980}]{PressureChangeActionPotentialMeasured:1980}
	\begin{barticle}
		\bauthor{\bsnm{{Iwasa, K and Tasaki, I. and Gibbons, RC.}}}:
		\batitle{Swelling of nerve fibers associated with action potentials}.
		\bjtitle{Science}
		\bvolume{17}(\bissue{4467}),
		\bfpage{338}--\blpage{9}
		(\byear{1980})
		\doiurl{10.1126/science.7423196}
	\end{barticle}
	\endbibitem
	
	\bibitem[\protect\citeauthoryear{Ling et~al.}{2020}]{NeuronalDeformation:2020}
	\begin{barticle}
		\bauthor{\bsnm{Ling}, \binits{T.}},
		\bauthor{\bsnm{Boyle}, \binits{K.}},
		\bauthor{\bsnm{Zuckerman}, \binits{V.}},
		\bauthor{\bsnm{Flores}, \binits{T.}},
		\bauthor{\bsnm{Ramakrishnan}, \binits{C.}},
		\bauthor{\bsnm{Deisseroth}, \binits{K.}},
		\bauthor{\bsnm{Palanker}, \binits{D.}}:
		\batitle{High-speed interferometric imaging reveals dynamics of neuronal
			deformation during the action potential.}
		\bjtitle{Proc Natl Acad Sci U S A}
		\bvolume{117}(\bissue{19}),
		\bfpage{10278}--\blpage{10285}
		(\byear{2020})
		\doiurl{10.1073/pnas.1920039117}
	\end{barticle}
	\endbibitem
	
	\bibitem[\protect\citeauthoryear{V{\'e}gh}{2025}]{VeghStokesEinstein:2025}
	\begin{botherref}
		\oauthor{\bsnm{V{\'e}gh}, \binits{J.}}:
		{How Hodgkin and Huxley measured that the axonal speed obeys the
			Stokes-Einstein relation}.
		Biosystems
		(2025)
	\end{botherref}
	\endbibitem
	
\end{thebibliography}

\end{document}